%% file: main.tex
\documentclass[acmsmall,screen]{acmart}

\input{style}

\input{macros}
\begin{document}
\input{titleAuthor.tex}

\maketitle

\input{introduction}

\input{program_synthesis}
\input{ltl}

\input{gpus}

\input{algo}

\input{lessons}

\newpage
\bibliographystyle{plain}
\bibliography{bib/semantic} 

\end{document}

%% file: style.tex
\usepackage{multicol}
\usepackage{xcolor}        
\usepackage{tcolorbox}     
\usepackage{listings}
\usepackage{makecell}
\usepackage{booktabs}
\usepackage{wrapfig}

\usepackage{amsmath} 
\usepackage{array}    
\usepackage{geometry} 

\definecolor{mygray}{rgb}{0.92,0.92,0.92}

\lstdefinelanguage{PythonCustom}{
    language=Python,                   
    morekeywords={parallel}, 
}

\lstdefinestyle{customPython}{
    language=PythonCustom,
    keywordstyle=\color{blue}\bfseries,
    commentstyle=\color{gray}\itshape,
    stringstyle=\color{red},
    basicstyle=\ttfamily\small,
    numbers=left,
    numberstyle=\tiny\color{gray},
    stepnumber=1,
    numbersep=5pt,
    backgroundcolor=\color{white},
    showspaces=false,
    showstringspaces=false,
    breaklines=true,
    frame=single,
    captionpos=b,
}

\lstdefinelanguage{CCustom}{
    language=C,                   
    morekeywords={parallel}, 
}

\lstdefinestyle{CPython}{
    language=CCustom,
    keywordstyle=\color{blue}\bfseries,
    commentstyle=\color{gray}\itshape,
    stringstyle=\color{red},
    basicstyle=\ttfamily\small,
    numbers=left,
    numberstyle=\tiny\color{gray},
    stepnumber=1,
    numbersep=5pt,
    backgroundcolor=\color{white},
    showspaces=false,
    showstringspaces=false,
    breaklines=true,
    frame=single,
    captionpos=b,
}

%% file: macros.tex
\newcommand{\NI}{\noindent}

\newcommand{\IMAGESIZE}[2]{\begin{center}\includegraphics[width=#1\textwidth]{images/#2}\end{center}}

{} 

\newcommand{\IE}{\textit{i.e.},\ } 
\newcommand{\EG}{e.g.,\ } 

\newcommand{\LTL}{{{\sc Ltl}}}                                                                                                                           
\newcommand{\LTLF}{{{\LTL$_{f}$}}}

\newcommand{\SUFFIXCLOSURE}[1]{\mathsf{sc}(#1)}

\newcommand{\CHARACTERISTIC}[2]{\mathbf{1}^{#1}_{#2}}
\newcommand{\CODE}[1]{\colorbox{mygray}{$\mathtt{#1}$}}
\newcommand{\SIZE}[1]{|\!|#1|\!|}

\newcommand{\AND}{\wedge}

\newcommand{\OR}{\vee}
\newcommand{\X}{\mathsf{X}}
\newcommand{\F}{\mathsf{F}}

\newcommand{\U}{\mathbin{\mathsf{U}}}

\newcommand{\EMPH}[1]{\emph{#1}}
\newcommand{\STRING}[1]{\langle #1 \rangle}
\newenvironment{GRAMMAR}{\[\begin{array}{lcl}}{\end{array}\]}

\newcommand{\VERTICAL}{\  \mid\hspace{-3.0pt}\mid \ }
\newcommand{\BOOL}{\mathbb{B}}
\usepackage{ifthen}
\newboolean{showcomments}
\setboolean{showcomments}{false}
\ifthenelse{\boolean{showcomments}}
  {\newcommand{\mynote}[2]{
    \fbox{\bfseries\sffamily\scriptsize#1}
    {\small$\blacktriangleright$\textsf{\emph{#2}}$\blacktriangleleft$}
   }
  }
  {\newcommand{\mynote}[2]{}
  }
\newcommand\NATH[1]{\mynote{Nath}{{\color{blue}#1}}}
\newcommand\MARTIN[1]{\mynote{Martin}{{\color{red}#1}}}

\newcommand{\IFTHENELSE}[3]{\CODE{if\; #1\; then\; #2\; else\; #3}}
\newcommand{\NOVSPACEPARAGRAPH}[1]{\NI\textbf{\emph{#1}.}}
\newcommand{\PARAGRAPH}[1]{\vspace{2mm}\NOVSPACEPARAGRAPH{#1}}
\newcommand{\FS}{\rightarrow}

%% file: titleAuthor.tex
\title{GPU accelerated program synthesis}
\subtitle{Enumerate semantics, not syntax!}

\author{Martin Berger}
\email{contact@martinfriedrichberger.net}
\orcid{https://orcid.org/0000-0003-3239-5812}
\affiliation{   
  \institution{University of Sussex and Montanarius Ltd}
  \city{London}
  \country{UK}
}

\author{Nathana\"el Fijalkow}
\email{nathanael.fijalkow@gmail.com}
\orcid{0000-0002-6576-4680}
\affiliation{
  \institution{CNRS, LaBRI and Universit\'e de Bordeaux}
  \city{Bordeaux}
  \country{France}
}

\author{Mojtaba Valizadeh}
\email{Valizadeh.Mojtaba@gmail.com}
\orcid{0000-0003-1582-3213}
\affiliation{
  \institution{Independent Researcher}
  \city{Cambridge}
  \country{UK}
}

%% file: introduction.tex
Program synthesis is an umbrella term for generating programs and
logical formulae from specifications.  With the remarkable performance
improvements that GPUs enable for deep learning, a natural question
arose: can we also implement a search-based program synthesiser on
GPUs to achieve similar performance improvements?  In this article we
discuss our insights on this question, based on recent
works~\cite{ValizadehM:ltlleaog,ValizadehM:seabasreioag}.  The goal is
to build a synthesiser running on GPUs which takes as input positive
and negative example traces and returns a logical formula accepting
the positive and rejecting the negative traces.  With GPU-friendly
programming techniques---using the semantics of formulae to minimise
data movement and reduce data-dependent branching---our synthesiser
scales to significantly larger synthesis problems, and operates much
faster than the previous CPU-based state-of-the-art.  We believe the
insights that make our approach GPU-friendly have wide potential for
enhancing the performance of other formal methods (FM) workloads.

%% file: program_synthesis.tex
\section*{Program synthesis}\label{program_synthesis}
Program synthesis \cite{DavidC:prosynsao,GulwaniS:prosyn} can be
classified along four orthogonal axes.
\begin{itemize}

\item What is the input (\IE how do we specify the desired result)?

\item What is the search space (\IE where are we searching)?

\item What is the search mechanism (\IE how do we traverse the search space)?

\item What is the output (\IE what do we synthesise)? 

\end{itemize}
The results of program synthesis are \EMPH{correct by construction},
assuming the program synthesiser is not buggy.
Correct-by-construction is desirable, and in marked contrast to
statistical learning algorithms.  Correct-by-construction is the prime
reason why program synthesis tools are used to generate synthetic
training data for deep learning\MARTIN{give reference}, and why
implementing performant program synthesis tools with deep learning as
search mechanism has so far not been reported in the open literature.
Program synthesis has two main flavours. They differ in the nature of
the input. In this paper, we focus on PBE (= programming by example),
where the input is a set of examples. For regular expressing synthesis
consider positive ($P$) and negative ($N$) examples as specification.
\begin{align*}\label{running_example_132}
  \begin{array}{lcl}
    \text{P} &:&
    \texttt{geon@ex.io} \quad
    \texttt{test@gmail.com} \quad
    \texttt{mail@test.org} \quad
    \texttt{mail@testing.com}
    \\
    \text{N} &:&
    \texttt{hello@} \quad
    \texttt{@test} \quad
    \texttt{email@gmail} \quad
    \texttt{t@test@gmail.com} \quad
    \texttt{mail with@space.com}
  \end{array}
\end{align*}
The synthesised regular expression should accept all positive examples and
reject all negative examples.  We expect that this simple
specification leads to the synthesis of a regular expression like
\begin{verbatim}
   [^@ \s]+@[^@ \s]+\.[^@ \s]+
\end{verbatim}
but there are infinitely many regular expressions that accept the
specification.  We (probably) also don't want the
synthesis simply to \EMPH{overfit} on the positive examples and return
\begin{center}
    \texttt{geon@ex.io}  \ |\ \ 
    \texttt{test@gmail.com}  \ |\ \  
    \texttt{mail@test.org}  \ |\ \ 
    \texttt{mail@testing.com}
\end{center}
To avoid this, program synthesis needs additional constraints,
typically the smallest (or at least small) program satisfying the
specification.  Alas, minimality makes synthesis hard
\cite{PittL:mincondpcbawap,GoldMD:comoautifgd,KearnsM:crylimolbfafa,AngluinD:onthecomlomiors}.
The trade-off between size and running time of the synthesiser is one
of the main parameters we have to affect the performance of program
synthesis.

\PARAGRAPH{A generic program synthesiser}
Program synthesis is often implemented by an endearingly simple and
generic algorithm: bottom-up enumeration, presented next as
\EMPH{dynamic programming} \cite{FislerK:datcenitc}:

\begin{lstlisting}[language=PythonCustom]
language_cache = atomic_predicates # global, memoises results

def enumerate(specification, size):
    for(op in operator_list): # for simplicity op has arity 2
        for(i + j = size - 1):
            for (left_program in language_cache(i)):
                for (right_program in language_cache(j)):
                    program = op(left_program,right_program)
                    if program satisfies specification:
                        return program
                    add_cache(program, language_cache(size))
\end{lstlisting}

\NI 
A call to \CODE{enumerate(specification, size)} generates all programs
of size \CODE{size}, assuming \CODE{language\_cache(i)} contains
all programs of size $i <$ \CODE{size}.  For each new candidate program we check
if it satisfies the specification.  If yes, we have found a solution
and terminate, otherwise we continue searching, but add the
new candidate to the language cache. 
The extreme simplicity of bottom-up enumeration makes it easy to
understand in great detail what the processor does, and that
makes performance benchmarking and optimisation more feasible.  This is
vital, as GPU performance can be extremely non-intuitive. 

Most previous research in program synthesis has been about
overcoming the scalability issues of bottom-up enumeration, usually
with heuristics, whether hard-coded or learned, for pruning the
search space.  Our research is radically different: we embrace the
exponential nature of enumeration, and exploit its simplicity in order
to make \EMPH{all cases} extremely fast on GPUs.  We do this by
noticing that the bottom-up enumeration algorithm above
\EMPH{sequentialises} the enumeration. This is not necessary!
Simplifying a great deal, all synthesis candidates can be generated
and checked for adherence to the specification in parallel. All
\CODE{for} loops in the code above, can be executed in parallel.  In
other words, bottom-up program synthesis is \EMPH{embarrassingly
  parallel}, making it a promising candidate for GPU-acceleration.

%% file: ltl.tex
\section*{Synthesising linear temporal logic formulae from examples}

Program verification means demonstrating that an implementation
exhibits the behaviour required by a specification. But where do
specifications come from?  Handcrafting specifications does not
scale. One possible approach is to learn them automatically from
example runs of a system. This is sometimes referred to as trace
analysis. A trace, in this context, is a sequence of events or states
captured during the execution of a system. Once captured, traces are
often converted into a form more suitable for further processing, such
as automata or logical formulae.

We explore the details of parallelising program synthesis by
converting traces to linear temporal logic (\LTL) \cite{PnueliA:temlogop}
as an example.  Having a concrete language lets us see clearly the
subtle interplay between the parallelism of the generic bottom-up enumeration
algorithm and the semantic details of the synthesised language that make
the synthesis task more GPU-friendly. \LTL\ is a modal logic, extending
propositional logic with connectives for specifying temporal
properties of traces, it is widely used in
industrial verification. The variant we use is
\LTLF\ \cite{DeGiacomoG:lintemlaldloft} which has the same formulae as
\LTL, but interpreted over finite traces. \EMPH{Formulae} over $\Sigma
= \{p_1, ..., p_n\}$ are given by the following grammar.
\begin{GRAMMAR}
  \phi
  &\quad ::= \quad\ &
  p_i \VERTICAL
  \neg \phi \VERTICAL
  \phi \AND \phi \VERTICAL
  \X \phi \VERTICAL
  \F \phi \VERTICAL
  \phi \U \phi
\end{GRAMMAR}
The atomic predicate $p_i$ means: at the current position $p_i$ is true.
The temporal connectives, $\X$ (for neXt), $\F$ (for Future) and $\U$
(for Until) are informally read as follows.
\begin{itemize}

\item $\X \ \phi$ means: at the next time step $\phi$ holds;
\item $\F \ \phi$ means: in the future $\phi$ holds;
\item $\phi\ \U \ \psi$ means: at some point $t$ in the future
     $\psi$ holds, but at all times before $t$, $\phi$ holds.

\end{itemize}
In \LTLF, we model the passing of time as a \EMPH{trace}, which is a
string using the powerset of $\Sigma$ as alphabet.  An example trace is
$\STRING{\{a\}, \{a,c\}, \emptyset}$, which we abbreviate to
$\STRING{a, ac, \emptyset}$.  If each position contains a set with one
element, for instance $\STRING{a, c, b}$, we write $acb$ for
readability.  For example, trace $\STRING{a, ac, \emptyset}$ models
that at time 0, only the atomic predicate $a$ holds, but $b$ and $c$
are false; at time 1, the atomic predicates $a$ and $c$ are true, but
$b$ is false; at time 2, all atomic predicates are false.  The
satisfaction relation is of the form $tr, i \models \phi$, where $tr$
is a trace and $i$ a natural number. It says that at time $i$, $\phi$
holds, assuming the passing of time is given by the trace $tr$.
Here are some examples.
\[
   squeegee, 0 \models \F g
      \qquad
   squeegee, 2 \not\models e
      \qquad
   squeegee, 2 \models \F e
      \qquad
   squeegee, 0 \models (\F g) \U \neg \F g
\]

\NI We now synthesise \LTLF\ formulae from examples to gain an
intuition for how synthesis works.  Let us call $(P, N)$ a
\EMPH{specification} when $P$ and $N$ are finite sets of example traces.  A
formula $\phi$ \EMPH{separates} $(P, N)$, if $tr, 0 \models \phi$ for
all traces $\phi \in P$, and $tr, 0 \not\models \phi$ for all traces
$\phi \in N$. We write $\phi \models (P, N)$ to signify that $\phi$
separates $(P, N)$.  Consider the specification $ P = \{\STRING{c},
\STRING{c,a}, \STRING{b}\}$ and $N = \{\STRING{b,a}, \STRING{a,a,c},
\STRING{b,a,c}\}$, a minimal separating formula is $\neg(b \U a)$ and
has size 4.  Here is a more complex example. 
\begin{itemize}

  \item Positive examples: $\STRING{ab, ab, a, b, a, ab}$,
    $\STRING{\emptyset, ab, a, b, b, \emptyset}$, $\STRING{ab, a,
    \emptyset, b, a, ab}$,  $\STRING{b, b, ab, b, a, ab}$, \linebreak$\STRING{ab,
    a, ab, a, \emptyset, a}$, $\STRING{a, ab, ab, b, a, ab}$,
    $\STRING{b, \emptyset, a, \emptyset, a, a}$

\item Negative examples:
$\STRING{b, ab, ab, b, a, ab}$, $\STRING{ab, ab, ab, a, \emptyset,
  a}$, $\STRING{\emptyset, ab, ab, b, a, a}$, $\STRING{b, b, a,
  \emptyset, a, ab}$, $\STRING{ab, ab, a, b, a, a}$, $\STRING{ab, a,
  b, b, a, a}$, $\STRING{b, a, ab, b, a, a}$.

\end{itemize}
After constructing 681,601,918 formulae, our GPU-based \LTLF-synthesiser
finds the solution
\[
    (\overline{b \AND \X\ b}) \U (a \U(\overline{ \X(\X(\overline{\X\ b}) \U a)}))
\]
in about 1 second (we write $\overline{\phi}$ for $\neg \phi$ for
readability).  This formula has size 16 and is of minimum cost by construction of our algorithm.

%% file: gpus.tex
\section*{Commandments of GPU-friendly programming}

\NATH{I propose removing this first paragraph, I feel that it's a bit less precise than the rest of the discussion, and I could only understand it properly after having read the rest, but then it doesn't add much information?}
Our journey towards implementing \LTLF-synthesis in a
GPU-friendly way starts with a core fact about
GPUs:  they have lots of parallel cores. Many more cores than
CPUs. This helps GPUs to deal with the cost of data movement,
sometimes referred to as the \EMPH{memory wall}.  The \EMPH{memory
  wall} refers to growing disparity between how quickly a processor
can read from memory, and the speed of \EMPH{compute
  instructions} (like addition or bit-shifting) that do not
access memory. According to \cite{DallyW:otmodocomp} in 2022:
\begin{itemize}

\item A 32 bit add instruction takes 150ps.
\item Moving the two 32 bit words to feed this addition 1mm takes 400ps.
\item Moving the 64 bits 40mm from corner to corner on a 400mm$^2$ chip takes 16ns.
\item Going off chip to fetch data for the addition takes 6ns per meter.

\end{itemize}

\NI The memory wall is ultimately rooted in the speed of electrons in
silicon, and, more importantly, in the number of gates the address has
to traverse on the path from the address generation unit in the processor,
all the way to the bits in memory (and back).  Digital gates transmit
information at most once per clock cycle.  We expect the memory wall
to become even worse in the future.

In order to help understand the performance characteristics of
processors that have to deal with the memory wall,
\cite{SamuelW:rooinsvpmfma} introduced the concept of \EMPH{compute
  intensity} (CI), which is a crisp quantity to optimise for when
programming GPUs.   The CI  of a processor is
defined as follows:
\[
   CI
      =
   \frac
      {\text{maximal number of machine instructions per second}}
      {\text{maximal number of bytes per second that can be moved     from memory to processor}}
\]
CI approximates from above the number of sequential compute
instructions a processor needs to do to avoid being stalled by memory.
Consider the Nvidia A100 GPU \cite{Nvidia:nvia100tcg}:
\begin{itemize}

\item 312 TeraFLOPs of BFLoat16 or FP16 compute instructions
  per second \MARTIN{FLOPs is about floating point ops, not general
    compute ops}

\item has a memory bandwidth of 2.039 TBytes per second.

\end{itemize}
Hence the compute intensity is $\frac{312}{2.039} = 153$. That means, assuming a
sequential processor, the programmer has to find 153 compute instructions
for every memory read to avoid stalling the processor.  That's a
pretty tall order: consider the C program below on the left. It is close to
something we use in our \LTLF\ synthesiser.

\begin{wrapfigure}{l}{0.3\textwidth}
\begin{lstlisting}[language=CCustom]
#include <stdint.h>

uint64_t f(uint64_t cs) {
    cs |= cs << 1;
    cs |= cs << 2;
    cs |= cs << 4;
    cs |= cs << 8;
    cs |= cs << 16;
    cs |= cs << 32;
    return cs;
}
\end{lstlisting}
\end{wrapfigure}

\NI If we compile it naively to RISC-V with Godbolt, we see code like
in the figure on the right generated. (See
\url{https://godbolt.org/z/6fz4oaYG9} for the full code.)

\begin{wrapfigure}{r}{0.4\textwidth}
  \IMAGESIZE{0.40}{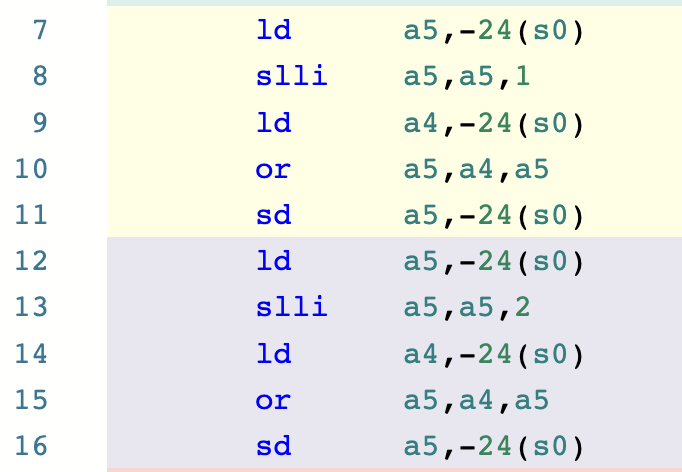}
\end{wrapfigure}

The details of the generated assembly do not matter. We only need to
know that \texttt{ld} (for ``load double word'') and \texttt{sd} (for
``store double word'') are data movement instructions.  Both transfer 64
bits between memory and processor. All other instructions are
compute instructions. We see that \texttt{slli} is directly
sandwiched between two loads, while \texttt{or} is followed by a store
and then another load. This falls well short of the CI of 153 that an
A100 needs (under naive sequential execution) to avoid stalling while
waiting for data to arrive from memory. GPU programmers can use \EMPH{massive
  parallelism} in order to get a higher compute intensity: whenever a
thread waits for a memory load, the GPU scheduler swaps it out for a
thread ready to compute. For this to be effective, we need several
things:

\begin{itemize}
  
  \item Zero-cost context switching
  \item The processor can support a large number of threads
  \item The application having enough parallelism

\end{itemize}
The first two are solved with modern GPUs.  The third is application
dependent and the programmer's responsibility. 
We have already seen the positive side of a combinatorial explosion: program synthesis by bottom-up enumeration has
an essentially \EMPH{unlimited supply} of parallelism in the
generation and checking of synthesis candidates. We shall now
investigate this parallelism in more detail and learn that it is
especially GPU-friendly.

\PARAGRAPH{Amortising the cost of data movement}
There is a second dimension to the cost of data movement: how can we
amortise it over the subsequent uses?  Consider matrix multiplication.
\begin{wrapfigure}{r}{0.3\textwidth}
  \IMAGESIZE{0.30}{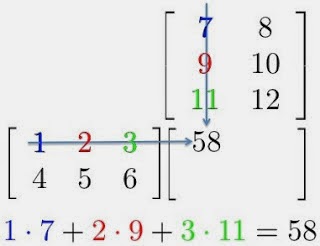}
\end{wrapfigure}
What happens to compute intensity of this algorithm as matrix gets
bigger?  Pay attention to number 7 in the image on the right. We load
it once and then compute $1 \cdot 7$, then $2 \cdot 7$, ..., $6 \cdot
7$ without ever reloading $7$ again.  In other words, we amortise the
cost of loading data from one matrix over all elements of the other
matrix. That means, \EMPH{the bigger the matrix, the higher the
  compute intensity}.  Matrix multiplication is one of the few
algorithms that has this property.  
This contributes to making deep learning so GPU friendly.

Let's come back to our naive bottom-up program synthesis algorithm.
Imagine we want to construct a formula  $\phi \AND \psi$
of
size $k$.  Let's
say we have constructed a formulae $\phi$ of cost $i$ and load it from
memory.  Let's assume the formulae with cost $k - i -1$ are $\psi_1,
\psi_2, ..., \psi_n$. Then the set of all relevant candidates that we
need to construct and check against the ambient specification $(P, N)$
is:
\[
   \phi \AND \psi_1 \qquad                              
   \phi \AND \psi_2 \qquad
       ... \qquad
   \phi \AND \psi_n
       \qquad\qquad\qquad\qquad\qquad\qquad\qquad\qquad
\]
In other words, the more synthesis candidates we have, the better we
can amortise the cost of loading $\phi$, \IE the better the
algorithm's compute intensity.  Since the number of synthesis
candidates increases dramatically as the synthesis problem gets
harder, we see another example of the positive side of computational
hardness.

\PARAGRAPH{SIMD and friends}
It's interesting to analyse the parallelism provided by bottom-up
enumeration: it is highly structured and, potentially,
GPU-friendly. In order to explain what that means, let us consider a
simplified CPU with four cores.

\IMAGESIZE{0.6}{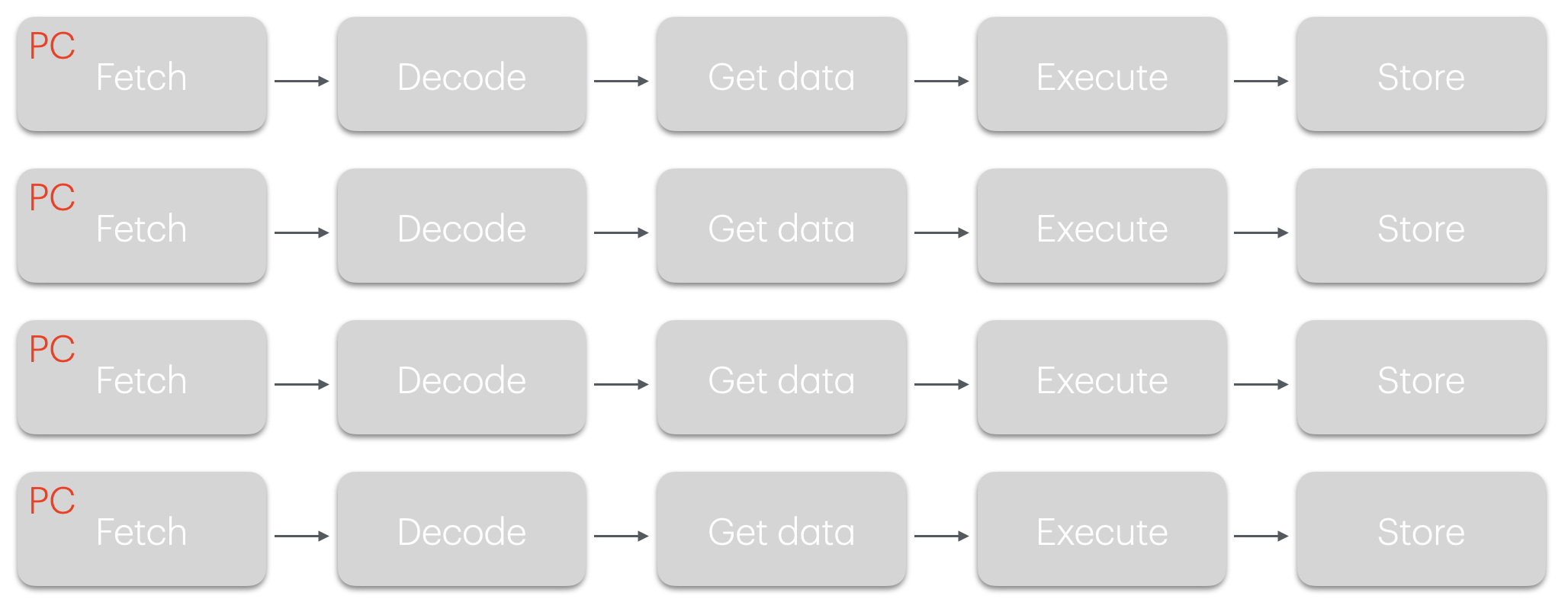}

\NI Each of the cores operates independently from the other
cores.  The price of this independence is that every core needs to
have, among other things, its own fetch and decode units. The fetch
unit retrieves the next command to be executed from memory (or an
instruction cache), while the decode unit then configures the rest of
the pipeline so it executes the command just retrieved by the fetch
unit. In a modern processor each fetch and each decode unit use a
large number of transistors, transistors that are not available for
other uses like addition or multiplication or bit-shifts. This
profligacy is unavoidable in general.  However, for some workloads,
the same program, or even exactly the same instructions, are executed
in lock-step on many different data items. Matrix multiplication is an
example: to compute $(A \cdot B)_{ij}$ we evaluate $\Sigma_{k} A_{ik}
\cdot B_{kj}$ for every element of the resulting matrix. In this case,
the fetch and decode units of a multi-core CPU would all redundantly
work on the same instructions.  In order to avoid such waste, the
\EMPH{single-instruction, multiple-data} (SIMD) architecture
\cite{flynnmj1972somecomputer,flynnMJ1966veryhighspeed}, and its many
variants, was invented.

\IMAGESIZE{0.6}{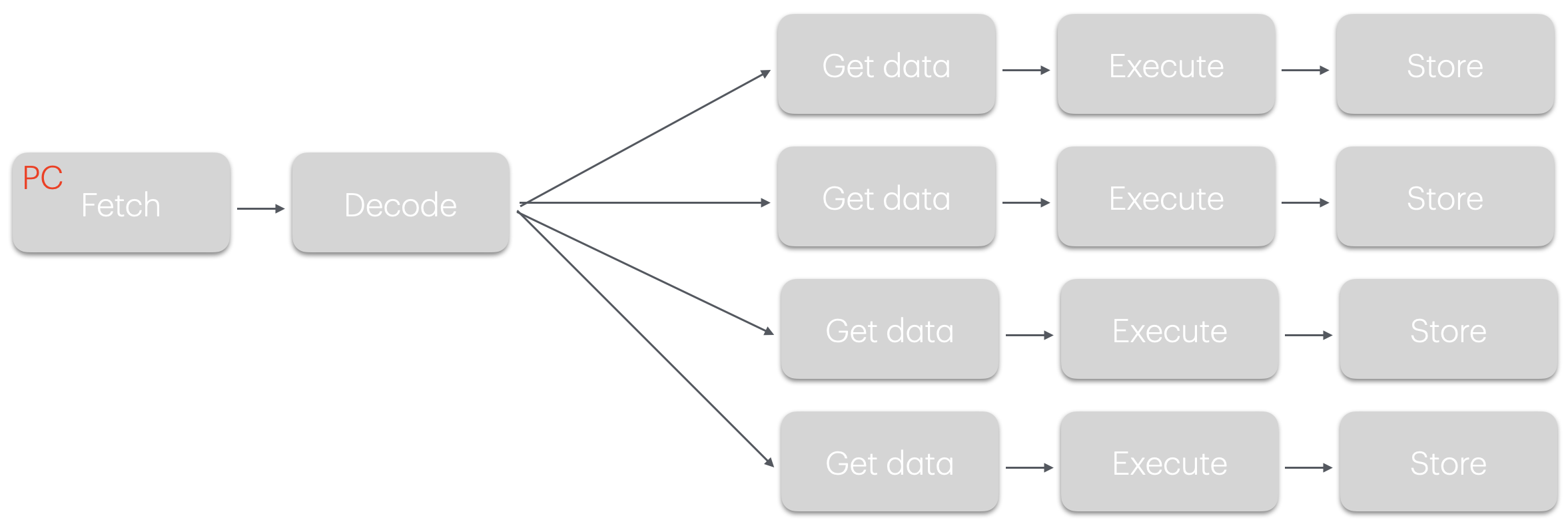}

\NI Here a single pair of fetch and decode units broadcast
instructions to all execute units, so they execute the same program,
indeed the same instructions, in lock-step.  The transistors saved are
used for more cores.  A subtle detail is that
symmetry-breaking is required to ensure that the different threads
execute the same instructions but work on different data.  The main
problem with SIMD is \EMPH{divergence}: what happens when the program
running on all cores does data-dependent branching, such as
\IFTHENELSE{x[threadID] > 15}{P}{Q}?  Some threads might need to run
\CODE{P} and others \CODE{Q}: they might need to run different
instructions!  There are numerous ways of dealing with this, including
dynamic warp formation
\cite{shoushtary2024controlflowmanagementmodern,FungWWL:dynwarfasfegcf},
but they still perform best when code is in a SIMD, lock-step
manner, so the programmer needs to minimise data-depended branching.


%% file: algo.tex
\section*{Parallel enumeration of \LTLF\ formulae on GPUs}

With a better understanding of the basic ideas behind GPU-friendly
programming under our belt, we can now delve into efficient synthesis
of \LTLF\ formulae.  The idealised program synthesis algorithm above
leaves many details open, in particular the \EMPH{exact} nature of the
search space. Let us focus on two burning questions:
\begin{itemize}

   \item How do we avoid the redundancies of syntax?
     Naively, we might think that the search space are
     \LTLF\ formulae, but formulae have a lot of redundancy, \EG
     $\phi \AND \psi$, and $\psi \AND \phi$ separate the same
     specifications $(P, N)$.

   \item Once we have decided on a search space, we need to consider
     minimising data movement, and that is strongly related to how we
     represent the search space in memory as bits.

\end{itemize}

\NI The first question is ultimately the most important because the
redundancies of syntax are infinite: for any $(P, N)$ there are
infinitely many \LTLF\ formulae separating $(P, N)$. We would like to
have a single representative for \LTLF\ formulae that separate the
given $(P, N)$ in the same way\MARTIN{What does ``in the same way''
  mean? Be more precise}.  The problem of canonical representatives in
program synthesis is so important that representing synthesis
candidates up to ``observational equivalence'' has become standard
terminology. 

Since the synthesiser only needs to check if a candidate formula
separates $(P, N)$, a natural candidate for the search space is
defined through the characteristic function $ \CHARACTERISTIC{}{\phi}
: (P \cup N) \FS \BOOL$: $ tr \mapsto 1$ iff $tr, 0 \models \phi$,
otherwise $ tr \mapsto 0$.
But there is a major complication with this search space, let us
illustrate it with an example.  Let's consider the specification $P =
\{squeegee\}, N = \emptyset$. If we were to use the (bitvector
representation of) functions $ (\{squeegee\} \cup \emptyset) \FS \BOOL
$ as search space, the bitvector had length 1.  Assume $bv$ was such a
bitvector, representing some formula $\phi$ that does not separate
$(\{squeegee\}, \emptyset)$. The naive bottom-up synthesis process
will eventually have to consider the (bitvector representation of the)
formula $\X\ \phi$. In other words, it will have to consider if $
squeegee, 0 \models \X\ \phi?  $ That is true, if and only if $
queegee, 0 \models \phi $.  Unfortunately, the bitvector $bv$ does not
contain any information about $queegee$ at all. So bitvector
representations of functions $(P \cup N) \FS \BOOL$ are insufficient
as search space.  The example hints at a solution, since $queegee$ is
an immediate suffix of $squeegee$.  We form the (non-empty)
suffix-closure (abbreviated $\SUFFIXCLOSURE{tr}$) of the trace and
order it by decreasing length, resulting in $(squeegee, queegee,
ueegee, eegee, egee, gee, ee, e)$.  More precisely, fix a total order
on $P \cup N$.
\begin{itemize}

\item A \EMPH{characteristic sequence (CS) for $\phi$ over $tr$} is a
  bitvector $cs$ such that $tr, j \models \phi$ iff $cs(j) = 1$. Note
  that that bits next to each other in $cs$, directly model the
  immediate suffix relation of subtraces of $tr$ in memory. We call
  this \EMPH{suffix-contiguity}.

\item A \EMPH{characteristic matrix (CM) representing $\phi$ over $P
  \cup N$}, is a sequence $cm$ of CSs, contiguous in memory, such
  that, if $tr$ is the $i$th trace in the order, then $cm(i)$ is the
  CS for $\phi$ over $tr$.

\end{itemize}
We have now arrived at our search space: the CMs associated with a
given specification $(P, N)$. 

\PARAGRAPH{Logical operations as bitwise operations}
Suffix-contiguity enables the efficient representation of logical
operations: if $cs = 10011$ represents $\phi$ over the word $abcaa$,
\EG $\phi$ is the atomic proposition $a$, then $\X \phi$ is $00110$,
\IE $cs$ shifted one to the left. Likewise $\neg \phi$ is represented
by $01100$, \IE bitwise negation.
Modern processors have instructions for bitshifts, $\X$ can be
executed as \EMPH{single} machine instructions (assuming the bitvector
is not too long)! In Python-like pseudo-code:

\begin{lstlisting}[language=Python]
def branchfree_X(cm):
    return [cs << 1 for cs in cm]
\end{lstlisting}

\NI Negation, conjunction and disjunction are equally efficient.  More
interesting is $\F$ which becomes the disjunction of shifts by
\EMPH{powers of two}, \IE the number of shifts is logarithmic in the
length of the trace (a naive implementation of $\F$ is linear).  For
clarity of presentation, we present the algorithm assuming the
characteristic matrix is 64 bits wide:

\begin{lstlisting}[language=Python]
def branchfree_F(cm):
    outCm = []
    for cs in cm:
        cs |= cs << 1
        cs |= cs << 2
        cs |= cs << 4
        cs |= cs << 8
        cs |= cs << 16
        cs |= cs << 32
        outCm.append(cs)
    return outCm

\end{lstlisting}

\NI
To see why this works, note that
$\F \phi$ can be seen as the infinite disjunction $\phi \OR \X\phi \OR
\X^2 \phi \OR \X^3 \phi \OR ...$, where $\X^{n} \phi$ is given by $X^0
\phi = \phi$ and $X^{n+1} \phi = \X \X^n\phi$.  Since we work with
finite traces, $tr, i \not\models \phi$ whenever $i \geq
\SIZE{tr}$. Hence checking $tr, 0 \models \F \phi$ for $tr$ of length
$n$ amounts to checking
\[
tr, 0 \models \phi \OR \X\phi  \OR \X^2 \phi \OR ...  \OR \X^{n-1}\phi
\]
The key insight is that the imperative update \CODE{cs \mathrel{|}= cs
  \ll j} propagates the bit stored at $cs(i+j)$ into $cs(i)$ without
removing it from $cs(i+j)$. Consider the flow of information stored in
$cs(n-1)$.  At the start, this information is only at index
$n-1$. This amounts to checking $tr, n-1 \models \X^{n-1}\phi$. Thus
assigning \CODE{cs\ \mathrel{|}= cs \ll 1} puts that information at
indices $n-2, n-1$. This amounts to checking $tr, 0 \models
\X^{n-2}\phi \OR \X^{n-1}\phi$.  Likewise, then assigning
\CODE{cs\ \mathrel{|}= cs \ll 2} puts that information at indices
$n-4, n-3, n-2, n-1$.  This amounts to checking $tr, 0 \models
\X^{n-4}\phi \OR \X^{n-3}\phi \OR \X^{n-2}\phi \OR \X^{n-1}\phi$, and
so on. In a logarithmic number of steps, we reach $tr, 0 \models \phi
\OR \X\phi \OR ...  \OR \X^{n-1}\phi$. This works uniformly for all
positions, not just $n-1$. In the limit, this saves an exponential
amount of work over naive shifting.  We implement $\U$ using similar
ideas, with the number of bitshifts also logarithmic in trace length.
In addition to saving work, exponential propagation maps directly to
machine instructions, and is essentially branch-free code for all
\LTLF connectives, thus maximises GPU-friendliness of our synthesiser.

\PARAGRAPH{Divide-and-conquer}
The synthesis presented so far does not attempt to split
specifications $(P, N)$ into smaller parts.  We ameliorate this
scalability limit by recursively splitting specifications until they
are small enough to be solved in one go, and then recombine the
results.  We could, for example, split $(P, N)$ into four smaller
specifications $(P_i, N_j)$ for $i, j = 1, 2$, such that $P$ is the
disjoint union of $P_1$ and $P_2$, and $N$ of $N_1$ and $N_2$. Then
$\phi_{ij}$ is synthesised recursively from the $(P_i, N_j)$. The
formula
$
  (\phi_{11} \wedge \phi_{12}) \vee (\phi_{21} \wedge \phi_{22})
$
separates $(P, N)$, but is not necessarily minimal: for example 
$\phi_{12}$ might already imply $\phi_{11}$.  We discuss some refinements
of this naive divide-and-conquer scheme in \cite{ValizadehM:ltlleaog}.

%% file: lessons.tex
\section*{Lessons for FM}

Program synthesis is not the only FM workload.  FM is a somewhat
loosely defined cluster of concepts and algorithms centring around
logic as core formalism, and aiming at making hardware and software
less buggy and more trustworthy. The main FM tools are SAT- and
SMT-solvers \cite{BiereA:handosat}, model-checkers
\cite{BaierC:priomodc}, and automatic theorem provers
\cite{BentkampA:mecmat}. The success in porting program synthesis to
GPUs raises the question: \EMPH{can we do the same for other FM
  workloads?} 
Recent positive results have been obtained along these lines for model counting~\cite{DBLP:journals/corr/abs-2006-15512}
and model checking~\cite{conf/birthday/Heemstra0W25,DBLP:conf/tacas/OsamaW24}.

At least conceptually, all FM algorithms search large search
spaces: SAT solvers the space of all assignment of boolean values to
propositional variables. First-order logic theorem proving search
proof-trees. Like program synthesis, they are of high computational
complexity: typically NP-complete or worse.  The scale and cost at
which we can do this search limits the scale of hardware and software
that humanity can trust. Hence there is a strong incentive to scale
them more!

A natural approach to acceleration of algorithms with high
computational complexity is \EMPH{parallelisation}.  Alas, despite
extensive research, including parallel SAT- and SMT-solving
competitions \cite{HeuleM:satcomtpt,BrombergerM:smtcomppact}, these
algorithms do not, today, seem to achieve achieve good performance on
GPUs.  In case of SAT solving, the community consensus seems to be
that \EMPH{``GPUs are hopeless''}.
If SAT-solving is indeed hopeless on GPUs, then SMT-solving is likely also hopeless, because SAT-solving
is a core building block of contemporary SMT-solving
algorithms. Moreover, \EMPH{bounded model-checking}
\cite{BiereA:boumodc}, one of the most widely used forms of
model-checking, also delegates much of the work to a SAT-solver.

So we are left with a conundrum: on the one hand, we have seen that at
least one FM workload benefits a great deal from GPU
acceleration. Moreover, future CPUs will be more and more GPU-like in
that they will have more and more SIMD units.  On the other
hand, experience so far suggests that GPUs are unlikely to help with
at least SAT-solving.

\PARAGRAPH{Pessimistic conclusion} The SAT experts are
  correct. Maybe SAT is an intrinsically sequential problem that does
  not benefit from parallelism, especially from the constrained
  (largely SIMD) parallelism that GPUs offer. Or maybe the memory
  access patterns that SAT-solvers require are too irregular, too
  unpredictable to be effectively maskable by GPUs' SIMD parallelism.
  Yet another possibility is that our success with GPU acceleration of
  program synthesis was a lucky fluke! After all, our synthesis is at
  the least expressive part of the Chomsky-Sch\"utzenberger hierarchy
  of languages: who knows if the refined bottom-up enumeration can
  still outrun the more severe combinatorial explosion of more
  expressive formalisms?

\PARAGRAPH{Wild optimism} On the other hand, there is also no proof at all that FM is
  intrinsically sequential.  Maybe the verification community worked
  so hard on making SAT-, SMT-solvers and model checkers fast on CPUs
  that they had no time to think much about parallelism.  The recent
  explosion in machine learning (ML) based on deep artificial neural
  networks, might give us pause for thought:
  \begin{quote}
    \EMPH{The biggest lesson that can be read from 70 years of AI research
      is that general methods that leverage computation are ultimately the
      most effective, and by a large margin. The ultimate reason for this
      is \EMPH{Moore's law}, or rather its generalisation of continued
      exponentially falling cost per unit of computation.}
    \cite{sutton2019bitter}
  \end{quote}
  Maybe we, the FM community, have been too smart. Let us follow the
  kids and leverage computation. The worst that can happen is that
  we fail!